\documentclass{optica-article}

\journal{opticajournal} 

\articletype{Research Article}
\usepackage{multirow}
\usepackage{pifont}
\usepackage{makecell}
\usepackage{color}
\usepackage{xspace}

\newcommand{\exvivo}{\textit{ex vivo}\xspace}

\newcommand{\invivo}{\textit{in vivo}\xspace}
\newcommand{\Invivo}{\textit{In vivo}\xspace}

\newcommand{\enfaceh}{\textit{en-face}\xspace}

\newcommand{\etal}{\textit{et al.\@}\xspace}

\newcommand{\um}{\(\mu\)m\xspace}

\newcommand{\ftx}[1]{\mathscr{F}_x\left[{#1}\right]\xspace}
\newcommand{\iftx}[1]{\mathscr{F}_x^{-1}\left[{#1}\right]\xspace}
\newcommand{\D}{\mathrm{D}\xspace}
\graphicspath{{.}{./figure}}

\begin{document}

\title{Multiple scattering suppression for \invivo optical coherence tomography measurement using B-scan-wise multi-focus averaging method}

\author{Yiqiang Zhu,\authormark{1} Lida Zhu,\authormark{1} Yiheng Lim,\authormark{1} Shuichi Makita,\authormark{1} Yu Guo,\authormark{1}and Yoshiaki Yasuno\authormark{1,*}}

\address{\authormark{1}Computational Optics Group, University of Tsukuba, Tsukuba, Ibaraki, Japan\\}

\email{\authormark{*}yoshiaki,yasuno@cog-labs.org}


\begin{abstract*}
We demonstrate a method that reduces the noise caused by multi-scattering (MS) photons in an \invivo optical coherence tomography image.
This method combines a specially designed image acquisition (i.e., optical coherence tomography scan) scheme and subsequent complex signal processing.
For the acquisition, multiple cross-sectional images (frames) are sequentially acquired while the depth position of the focus is altered for each frame by an electrically tunable lens.
In the signal processing, the frames are numerically defocus-corrected, and complex averaged.
Because of the inconsistency in the MS-photon trajectories among the different electrically tunable lens-induced defocus, this averaging reduces the MS signal.
Unlike the previously demonstrated volume-wise multi-focus averaging method, our approach requires the sample to remain stable for only a brief period, approximately 70 ms, thus making it compatible with \invivo imaging.
This method was validated using a scattering phantom and \invivo unanesthetized small fish samples, and was found to reduce MS noise even for unanesthetized \invivo measurement.   
\end{abstract*}


\section{Introduction}
Optical coherence tomography (OCT) is a noninvasive imaging technique with a resolution of a few to tens of micrometers, and has been used for clinical imaging in fields such as ophthalmology\cite{Geitzenauer2011BJO, Geevarghese2021ARVS, Ang2018PRER, Venkateswaran2018EV} and cardiology\cite{yonetsu2013optical,vignali2014research}.
Because of its long imaging depth, OCT recently has been adopted for noninvasive and nondestructive microscopy \cite{chen2012review}.
OCT can visualize the deep tissue region at around several hundred of micrometers to a few millimeters from the sample surface, and it is significantly larger than that of conventional microscopy, which is only around a few tens to hundreds of micrometers.

Conventionally, the imaging depth of OCT was believed to be dominated by two factors, the depth of focus of the probe optics and the scattering of the sample.
To overcome the former limit, several methods have been successful.
For instance, the fusion of multiple images with several depth focus positions \cite{Drexler1999OL}, computational refocusing methods \cite{Coquoz2017OpEx, yasuno2006non, Ralston2007NatPhys, Kumar2014OpEx}, and the combination of complex signal processing and focus fusion such as Gabor-domain OCT \cite{Rolland2010OpEx, Yoon2019BOE}.
To overcome the tissue-scattering limit of the imaging depth, longer wavelength probe has been applied.
In general, a 1.3-\um wavelength probe has a better image penetration than a 830-nm or visible OCT.
For retinal imaging, a 1.0-\um probe was shown to have a higher penetration \cite{Unterhuber2005OpEx, Povazay2003OpEx, Lee2006OpEx, Yasuno2007OpEx}, and it has become a common probe wavelength band for clinical retinal imaging.
More recently, even longer wavelength, such as 1.7 \um, has been used to investigate samples with high scattering, such as cardiovascular tissues \cite{li2017intravascular} and brain tissues \cite{zhu20211700}, and have demonstrated higher penetration than OCT using shorter wavelengths, such as more than 3 mm in coronary arterial tissues or around 1.3 to 1.6 mm in brain tissues.

Because these two limiting factors of imaging depth have been overcome, the multiple scattering (MS) has gradually been recognized as an additional limiting factor.
In general, OCT imaging theory is based on the assumption that most MS photons are rejected by a confocal pinhole (i.e., a single-mode fiber tip) \cite{yao1999monte} and the single-scattering (SS) photons govern the imaging.
In practice, however, some MS photons are captured through the confocal pinhole and appear in the image.
Because an MS photon has a longer optical path than that of the SS photons, a photon that undergoes multiple scattering at a specific depth appears in the image to be located at a deeper location.
In cotrast, the contribution of SS photons becomes less at the deeper depth in the image.
Hence, deeper regions in an OCT image are more dominated by MS photons.
This dominance of MS photon degrades the resolution and contrast of the OCT image in the deep regions \cite{wang2002signal}.
In addition, it degrades the quantitative measurement capability of functional OCT, such as polarization sensitive OCT \cite{zhu2024polarization}.

Several methods have been used to mitigate the MS effect.
For example, Badon \etal proposed the smart OCT method, which modulates the pupil using a spatial light modulator (SLM) and pre-defined reflection matrix\cite{badon2016smart}. 
Borycki and associates used SLM-based pupil modulation and spatial correlation theory to reduce the MS effect (or equally, the coherent cross-talk) of full-field swept-source OCT\cite{Borycki2019BOE, wojtkowski2019spatio, Auksorius2022iScience}.
Liu \etal proposed the aberration-diverse method for MS suppression\cite{liu2018aberration}.
In this method, intentional astigmatism was introduced using a deformable mirror, and multiple, typically twelve, OCT volumes were acquired with different astigmatism axes.
After correcting the astigmatism using computational adaptive optics \cite{adie2012computational}, the volumes were coherently averaged.
The paths of the MS photons are not consistent across the different astigmatism axes, whereas those of the SS photons are consistent.
Hence, the coherent average can reduce the MS-photon contributions.

Although the aforementioned modalities have successfully mitigated the MS effect, they require expensive wavefront manipulation devices, such as an SLM or deformable mirror.
We previously proposed the multi-focus-averaging (MFA) method, which is based on an approach similar to that of the aberration diverse method, but we used a cost-effective electrical tunable lens (ETL) \cite{zhu2023multi}.
Multiple, typically seven, OCT volumes are acquired with different defocus positions, and the volumes are coherently averaged after correcting the defocus using computational defocus correction.
This method was also applied to Jones-matrix based PS-OCT (JM-OCT), and mitigation of polarization artifacts was demonstrated \cite{zhu2024polarization}.

Although the aberration diverse method and MFA perform well on static samples, such as static phantom and postmortem samples, its application to \invivo measurement still poses a great challenge.
Because these methods rely on coherent (i.e., complex) averaging of multiple volumes, the phases of these volumes should be consistent.
In other words, the sample should be highly stable during the multiple volumetric acquisition, which usually takes a few tens of seconds.

In this work, we propose a new version of the MFA method, referred to as the B-scan-wise-MFA (B-MFA) method for MS suppression in \invivo measurement.
This new method sequentially acquires multiple cross-sectional OCT frames, instead of volumes, with different defocus.
The defocus of each frame is corrected by applying a one-dimensional (1-D) version of computational refocusing, and all defocus-corrected frames are coherently averaged to reduce the MS signals.
Note that throughout the manuscript, a single cross-sectional scan is denoted as a frame, while a B-scan refers to a set of frames acquired at the same location.
This method requires phase stability only during the acquisition time of a few frames (i.e., a B-scan), not of volumes, so the required stable time is typically less than 100 ms.
Hence, this method is applicable to \invivo measurement.
The performance of the B-MFA method was validated by measuring a scattering phantom and \invivo small fish.
We also discuss the optimization of parameters for the B-MFA method, the artifacts related to 1-D computational refocusing, and their impact on the final images.


\section{Principle and implementation of B-MFA method}
\subsection{Data acquisition and signal-processing flow}
\begin{figure}
	\centering\includegraphics{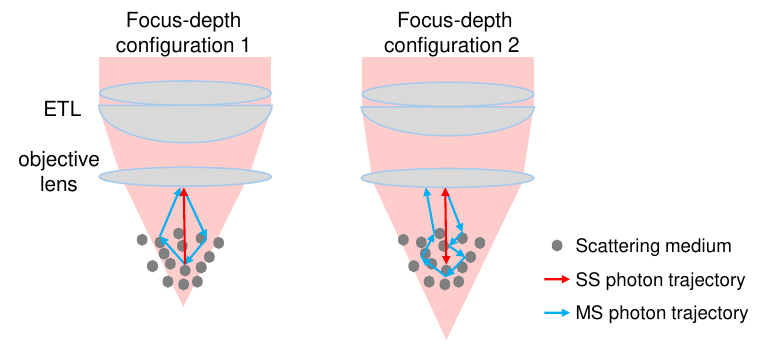}
	\caption{
		Example trajectories of single-scattering (SS) and multi-scattering (MS) photons with two focus configurations of an electrically tunable lens (ETL). 
        The SS photon was scattered only once, and hence its trajectory is consistent regardless of the ETL configuration.
        In contrast, the trajectories of MS photons become inconsistent when the ETL configuration altered.
	}
	\label{fig:trajectory}
\end{figure}
Before describing the details of the B-MFA method, we first present an overview of this method.
The B-MFA method is based on the assumption that the trajectories of MS photons are not consistent if the depth positions of the focus are different\cite{zhu2023multi}, as depicted in Fig.\@ \ref{fig:trajectory}.
Hence, the first step of B-MFA is to acquire multiple frames with different focus positions using an ETL.
The details of the data acquisition protocol are described in Section \ref{sec:scanProtocol}.
In the second step, we correct the defocus using a 1-D phase-only spatial frequency filter  (Section \ref{sec:refocusing}).
Finally, the frames are complex averaged after the axial shifts and phase offsets of the frames have been corrected.
Because we used a JM-OCT in our particular implementation, we additionally correct the bulk-phase offset in the four polarization channels of the JM-OCT. 
The details of the shift and phase corrections as well as the complex-averaging process are described in Section \ref{sec:complexAverage}

Because MS-photons trajectories are different in frames with different focus depths, the randomized MS signal is reduced by the complex averaging, whereas the defocus corrected SS-signal is not.
In this manuscript, we refer to this complex-averaged image as a ``B-MFA image.''
For volumetric measurement, we sequentially acquire cross-sectional B-MFA images at several slow-scan positions.

\subsection{Data acquisition protocol}
\label{sec:scanProtocol}
\begin{figure}
\centering\includegraphics{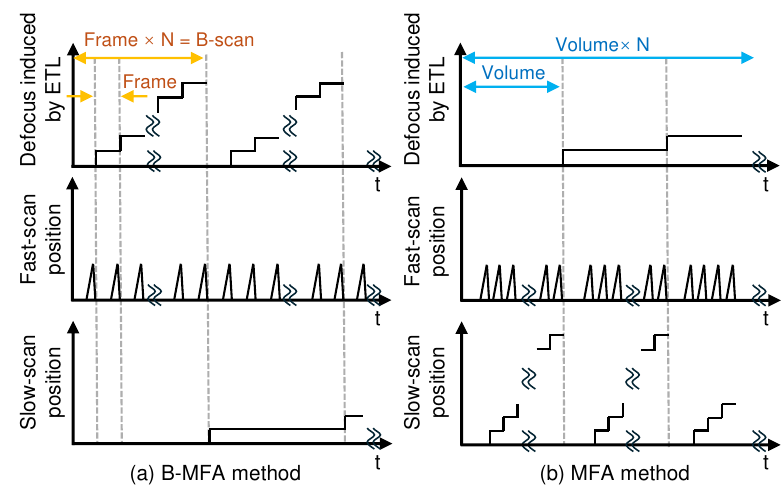}
\caption{
	Scanning protocol for the B-MFA method(a) and previously proposed MFA method \cite{zhu2023multi} (b).
	N refers to the number of repeated acquisitions (i.e., frames) which are complex averaged after defocus correction to reduce the MS signals.
	For both methods, the defocus is altered by an ETL.
	For B-MFA, the defocus is altered for each frame, whereas for MFA, the defocus is altered for each volume.
	}
	\label{fig:scanProtocol}
\end{figure}
The first step of the B-MFA imaging is to acquire multiple frames with different depth positions of focus.
Here the depth positions of the focus are actively controlled by an ETL that is equipped in the sample arm of the OCT \cite{zhu2023multi}.
The implementation details of the OCT system are described later in Section \ref{sec:JMOCT}.

The data acquisition protocol of B-MFA is summarized in a schematic time chart in Fig.\@ \ref{fig:scanProtocol}(a).
As shown in this diagram, multiple frames are sequentially acquired at each slow-scan (i.e., B-scan) position.
For every frame acquisition, the ETL updates the depth position of the focus so that all frames are acquired using different focus positions.
This measurement process is repeated for each slow-scan location to obtain a volumetric dataset.

In our typical implementation, a single frame acquisition time including the focus transition time of ETL is 13.4 ms and the number of frames at a single slow-scan location (i.e., the number of frames per B-scan) is five (see Section \ref{sec:measurementProtocol} for details).
Hence, the typical acquisition time for a single slow-scan location is 67 ms.

\subsection{Computational refocusing}
\label{sec:refocusing}
\subsubsection{Computational refocusing using a 1-D phase filter}
\label{sec:1dRefocusing}
After data acquisition, each frame is then processed for 1-D computational refocusing.
Here, the 1-D lateral complex signal at each depth of the frame is processed by a 1-D phase-only spatial frequency filter designed based on the Fresnel-diffraction model \cite{yasuno2006non}.
For a defocus distance (i.e., the distance from the focus to the imaging depth) of $z_{d}$, the phase only filter is
\begin{equation}
H^{-1}\left(f_{x};z_{d}\right) =
 \exp\left(\frac{-i\pi\lambda_{c}z_{d}f_x^2}{2}\right),
\end{equation}
where $f_{x}$ denotes the spatial frequencies corresponding to the fast-scan lateral position $x$, and $\lambda_{c}$ is the center wavelength of the probe beam.

The refocused frame $S'(x;z)$ is obtained using this filter and two sequential 1-D Fourier transform operations as
\begin{equation}
S'(x;z) =
 \iftx{ \ftx{S(x;z)}H^{-1}\left(f_{x};z_{d}(z) \right)},
\end{equation}
where $S(x;z)$ is the original complex frame and $\ftx{\quad}$ and $\iftx{\quad}$ are the 1-D Fourier transform and its inverse Fourier transform, respectively, along the fast-scan ($x$) direction.
Here, $z_{d}$ is considered to be a function of the depth in image $z$. 
In our implementation, $z_d$ is estimated from the measured data, as described in detail in the next section (Section \ref{sec:defocusEstimation}). 

Note that this method corrects the defocus only along the fast-scan direction. 
The impact of this limitation is discussed in Section \ref{sec:limitOf1D}.

\subsubsection{Estimation of the defocus distance}
\label{sec:defocusEstimation}
In our method, the defocus distance $z_{d}$ is estimated from the measured OCT images, where the information entropy of the OCT images is used as a sharpness metric.
Note that here, we use the information entropy of an 2D \enfaceh OCT image at each depth, although the refocusing was performed for each lateral ($x$-) line individually.
This is because a 1-D lateral signal is not informative enough to compute a reliable sharpness metric.

The defocus-distance estimation is performed at each depth, and hence these initial estimates are obtained as a function of depth.
We then extract the estimates from the depth region in which the estimated focus distances are linear to the depth, and use them to estimate the defocus distances over the range of all depths in the image.
Specifically, the estimated defocus distance is linear-fit to the depth using an intensity-weighted linear regression.
This linearly fitted line gives the final estimates of the defocus distance throughout the whole depth range.

\subsection{Shift and phase-offset corrections and complex averaging in JM-OCT}
\label{sec:complexAverage}
In our implementation, we used JM-OCT, which acquires four OCT cross-sectional images at each slow-scan position\cite{yasuno2023multi} that corresponds to the four polarization entries of the Jones matrix.
Here, we describe the methods to correct the mutual phase offsets in the four images in addition to the shift and phase-offset corrections in the multiple frame acquisitions because they are crucial to complex averaging the refocused OCT signals.

A Jones matrix cross-sectional image (JM cross-section) consists of four complex OCT images corresponding to the four polarization channels.
For computational refocusing, we estimate the defocus distance using only one polarization channel with the method described in Section \ref{sec:defocusEstimation}, and we apply it to all polarization channels.
Namely, the defocuses of all the polarization channels are corrected with the same estimated defocus distance.

\begin{figure}
	\centering\includegraphics[width=10cm]{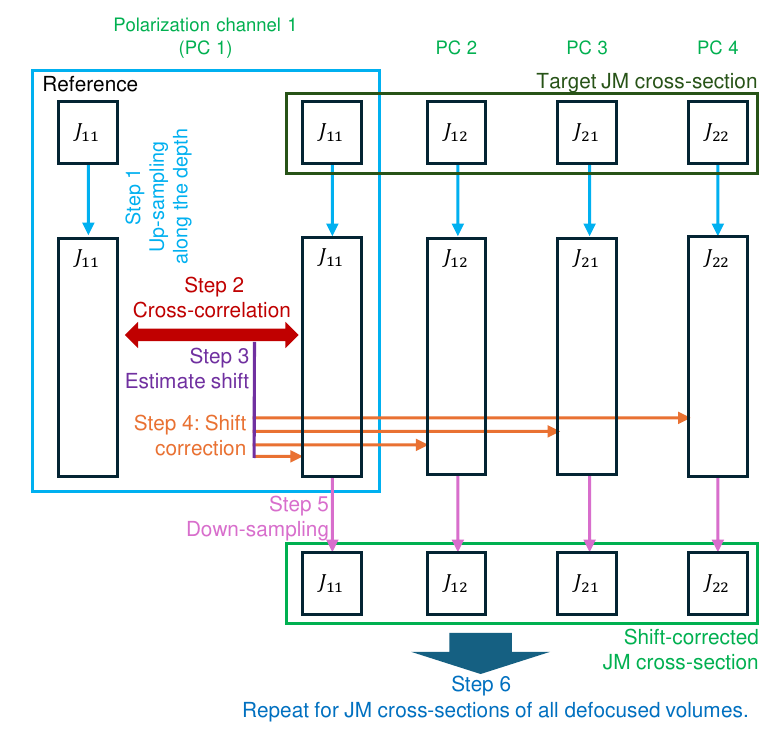}
	\caption{
		The processing diagram of axial-shift correction. The axial shift mainly induced by the modulation of the ETL is detected by using the first polarization channel (PC 1, blue box). 
		For this detection, the cross-sectional image is up-sampled along the depth (step 1). 
		The reference and target images are cross-correlated (step 2), and the shift amount is estimated (step 3). 
		This estimated shift amount is used to correct the shifts of all polarization channels (step 4). 
		Finally, the images are down-sampled to form a shift-corrected JM cross-section (step 5). 
		This process is repeated for all JM cross-sections of all defocused volumes.
	}
	\label{fig:ShiftCorrection}
\end{figure}
Because of the deformation of the ETL, there are non-negligible depth shifts among the images taken with different defocus. 
We corrected the depth shifts as summarized in the diagram presented in Fig.\@ \ref{fig:ShiftCorrection}. 
We compute the shift using the cross-correlation function of linear-intensity OCT images (blue box in the diagram), where the linear intensity images are obtained from a single polarization channel and the intensity image of the first defocus value is used as the reference. 
Before computing the cross-correlation function (Step 2 in the diagram), the images are up-sampled four times along the depth using Fourier-domain zero-padding to achieve sub-pixel accuracy (Step 1). 
The cross-correlation function is computed along the depth by a direct method (i.e., not the Fourier-domain method). 
The amount of shift is determined from the peak of the cross correlation function (Step 3), and the shifts of all polarization channels are corrected by using this estimated shift amount (Step 4). 
After correcting the shift in the up-sampled images, the images are down-sampled to the original pixel resolution using Fourier-domain de-padding (Step 5). 
And finally, this process is repeated for all JM cross-sections of all defocused volumes. 

\begin{figure}	
	\centering\includegraphics[width=11cm]{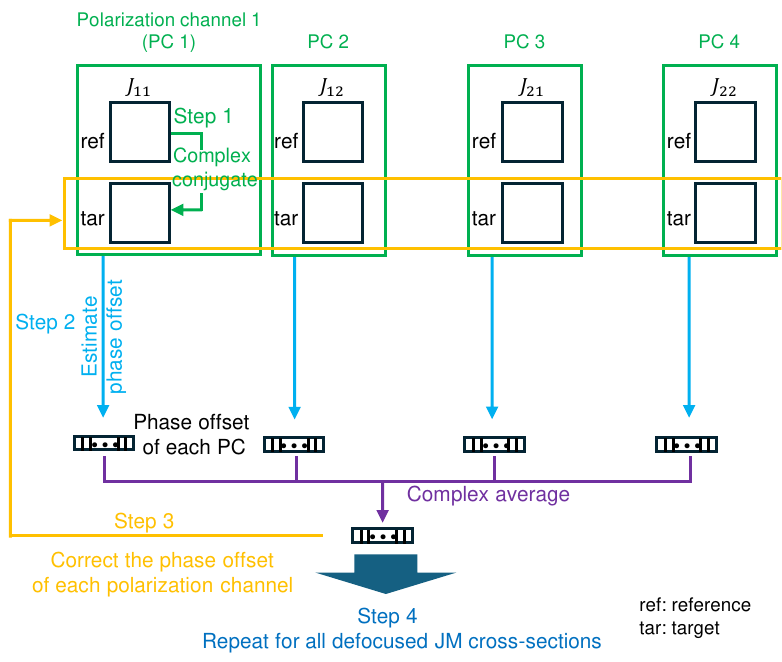}
	\caption{
		Flow of phase-offset corrections.
		The phase offsets in the images obtained with different defocus are corrected for each polarization channel.
		We first compute the product of the target (tar) image and the complex conjugate of the reference (ref) image (step 1, green boxes).
		Each product image is complex averaged along the depth for each polarization channel (step 2, light-blue arrow), and the resultant 1-D arrays are further complex averaged over the polarization channels (purple arrows).
		The phase of the resulting 1-D complex array represents the phase offset of each A-line.
		The phase offset of each A-line in the target image is then corrected (step 3, yellow arrow and box).
		These processes are repeated for all JM cross-sections (step 4).
	}
	\label{fig:phaseCorrection}
\end{figure}
After correcting the axial shifts, the mutual phase offsets in the images with different defocus are estimated and corrected.
Here, we consider the phase offsets in the JM cross-sections, where a JM cross-section is a set of four complex OCT images one for each of the four polarization channels (PCs).
We first estimate the phase offsets of the images at each PC independently.
For this estimation, we use only the depth region of 30 pixels from the sample surface (see Appendix \ref{sec:segmentation} for details of the sample surface detection).
The first JM cross-section is used as a reference, and the phase offsets between the complex images of the reference JM cross-section and the images of the target JM cross-section are estimated.
The complex image of each PC in the target JM cross-section is multiplied by the complex conjugate of the corresponding PC's complex image in the reference JM cross-section (Step 1 in Fig.\@ \ref{fig:phaseCorrection}), and then complex averaged along the depth (Step 2).
Because this operation is performed for each of four PCs, four 1-D complex arrays along the fast scan direction are given (the phase offset of each PC in the figure).
By averaging these four 1-D complex arrays corresponding to the PCs and taking the phase, we obtain a 1-D array of the phase offsets where each entry of the array represents the phase offset of each A-line.
Finally, the mutual phase shift of the reference JM cross-section and the target JM cross-section is corrected for all depths by subtracting the estimated phase offsets from the complex images in the target JM cross-section, i.e., multiplying the conjugate phase in complex (Step 3).
This operation is repeated for all frames (i.e., JM cross-sections) so that the phase offsets of all the frames are corrected (Step 4).

After the phase offset corrections, the OCT images of each PC are complex averaged over the frames to generate four MS-reduced complex OCT images, which forms a MS-reduced JM cross-section.
The final B-MFA image is obtained by averaging the squared intensities of four complex images of the MS-reduced JM cross-section.

Note that the aforementioned operations are for one slow-scan position, i.e., B-scan position.
We repeat these operations for all other slow-scan positions to obtain a B-MFA volume.

\section{Validation design}

\subsection{JM-OCT setup}
\label{sec:JMOCT}
A custom-built passive-polarization-delay (PPD) based JM-OCT was used to evaluate the B-MFA method.
A PPD module splits the probe beam into two orthogonal polarizations and applies different delays to them.
In addition to this probe-beam polarization multiplexing, polarization-diversity detection is used.
Hence, four complex OCT images corresponding to the four polarization channels (i.e., two multiplexed polarizations of the probe beam times two detection polarizations) are obtained.

The center wavelength and scanning bandwidth of the light source (AXP50124-8, AXSUN, MA, USA) are 1,310 nm and 106 nm, respectively.
The effective focal length of the objective lens (LMS03, Thorlabs, NJ, USA) used in the system is 36 mm.
The lateral and axial resolutions are 17 $\mu$m (in $1/e^2$-width) and 14 $\mu$m (in full-width-half-maximum) in tissue, respectively.
The depth-of-focus (DOF) without the ETL is 0.36 mm.
The A-line rate of the system is 50,000 A-lines/s.
More details of the JM-OCT principle \cite{yasuno2023multi} and the implementation of the particular JM-OCT system used in this study \cite{miyazawa2019polarization,li2017three} can be found in elsewhere.

An ETL (EL-10-30, CI-NIR-LD-MV, Optotune, Switzerland) is used in the sample arm to axially shift the focus position.
The details of the sample arm equipped with the ETL are described in \cite{zhu2023multi}.
As we investigated in the previous study, the focus modulation of the ETL alters the lateral resolution only by a negligible amount, i.e., less than 0.5 $\mu$m (Section 4.3.4 of Ref.\@ \cite{zhu2023multi}).

Note that, although we used a JM-OCT in this study, the B-MFA method can be applied to standard (i.e., non-polarization-sensitive) OCT.

\subsection{Measurement protocol}
\label{sec:measurementProtocol}


At each slow-scan (i.e., B-scan) position, five continuous cross-sectional frames were acquired as the defocus was incremented 0.18 mm at each acquisition.
The defocus increment was equivalent to half the DOF, where the DOF is that without the ETL.
The total focus shift of five frames was 0.72 mm.
These parameters were determined by an optimization experiment that is described in details in Section \ref{sec:optimization}.
Each cross-sectional frame consists of 256 A-lines, and the acquisition time of a single frame was 5.12 ms.
By accounting for the defocus transition time of ETL (7.5 ms) and the pullback time of the galvanometer scanner (0.8 ms), the five continuous frames were acquired in 67.1 ms.
The phase should be stable during this acquisition time.
The acquisition was repeated for 256 slow-scan positions, and the total time to acquire a volume was 17.18 s.
In summary, the volumetric acquisition time and required phase-stable duration were 17.18 s and 67.1 ms, respectively.

For reference, we acquired or generated three additional volumetric images. 
The first image is a single acquisition image, which was made by extracting only the third frame of the five sequential frames acquired for B-MFA. 
Although no averaging was performed, the defocus was corrected.

The second reference is the single frame averaging (SFA) image.
Here, we acquired a volume following the B-MFA protocol but without shifting the focus.
This volume was processed in a manner identical to that of the B-MFA, i.e., the defocus correction, shift and phase corrections, and complex averaging were all the same.

The third reference is a standard MFA image \cite{zhu2023multi}.
Here, five OCT volumes were sequentially acquired with different defocus, as shown in Fig.\@ \ref{fig:scanProtocol}(b).
The increments in defocus between two consecutive volume acquisitions were the same as that of the B-MFA.
Each volume was acquired with a standard raster scan with 256 $\times$ 256 A-lines, and the total acquisition time of the sequential five volumes was 9.92 s.
For MFA, the phase should be stable during the five volume acquisitions, and hence it was 9.92 s, which is 148 times longer than the required phase-stable time of B-MFA.

For all scan protocols, the lateral scanning range was 1.5 mm $\times$ 1.5 mm.
The lateral field was covered with 256 $\times$ 256 lateral sampling points, which yielded a lateral pixel separation of 5.86 \um that is around 1/3 of the lateral resolution.

\subsection{Samples}
\label{sec:samples}
\begin{figure}
	\centering\includegraphics{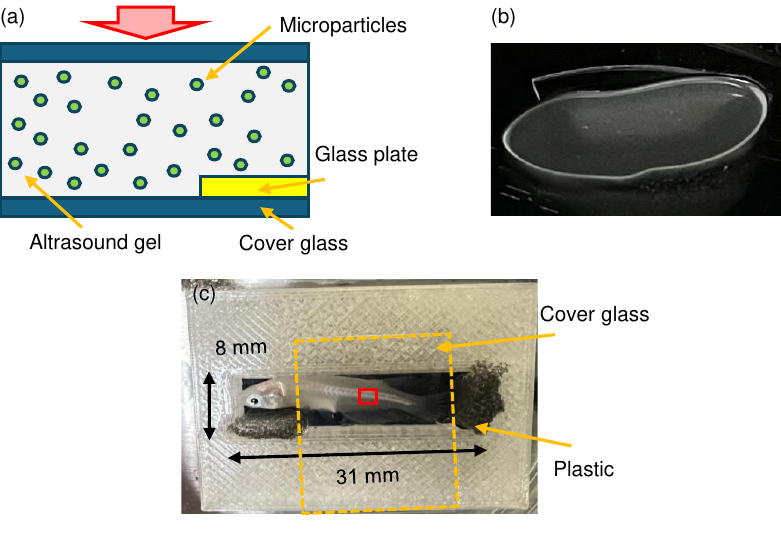}
	\caption{%
		Schematic (a) and photograph (b) of the scattering phantom.
		The phantom consists of a scattering layer, which is a mixture of polystyrene micro-particles and ultrasound gel, and a glass plate embedded in the scattering layer.
		The glass plate is used to provide a scattering-free area.
		The red arrow in (a) indicates the direction from which the photograph of (b) was taken.
		(c) An \invivo medaka sample held in a 3-D printed container.
		The container was filled with water and fish was not anesthetized.
		A thin cover glass was put on the container to prevent the fish from accidentally jumping out.
		The red box roughly indicates the measured area of the OCT.
	}
	\label{fig:sample}
\end{figure}
A scattering phantom and ten \invivo medaka fish were measured to evaluate the proposed method.
The scattering phantom consists of four parts, two cover glasses, a scattering layer, and a glass plate under the scattering layer that creates a space without a back scattering signal.
The scattering layer is a mixture of 0.04-mL polystyrene microparticles (diameter of 10 \um, 72986-10ML-F, Sigma-Aldrich, MO, USA) and 0.5-mL ultrasound gel (pro Jelly, Jex, Japan).
This mixing ratio results in a particle concentration of 13.5 $\times$ 10$^3$ particles/mm$^3$.
The schematic and photograph of the phantom are shown in Fig.\@ \ref{fig:sample}(a) and (b).

The medaka, also known as the Japanese rice fish, is a small fish similar in size to a zebrafish and is widely used as a model animal in biological research.
We measured ten \invivo medakas without anesthesia.
A fish was placed in a 3-D printed container with a water-filled groove that was 5 mm $\times$ 8 mm $\times$ 31 mm (height $\times$ width $\times$ length) in size.
We placed a thin glass slip above the container to prevent the fish from accidentally jumping out during the measurement.
Figure \ref{fig:sample}(c) shows a photograph of the container and a sample.
In this figure, the red box roughly indicates the scan area.

The protocol of the fish experiment follows the animal experiment guidelines of the University of Tsukuba and is approved by the Institutional Animal Care and Use Committee of the University of Tsukuba. 

\section{Result}
\label{sec:allresults}
\subsection{Scattering phantom}
\begin{figure}
	\centering\includegraphics{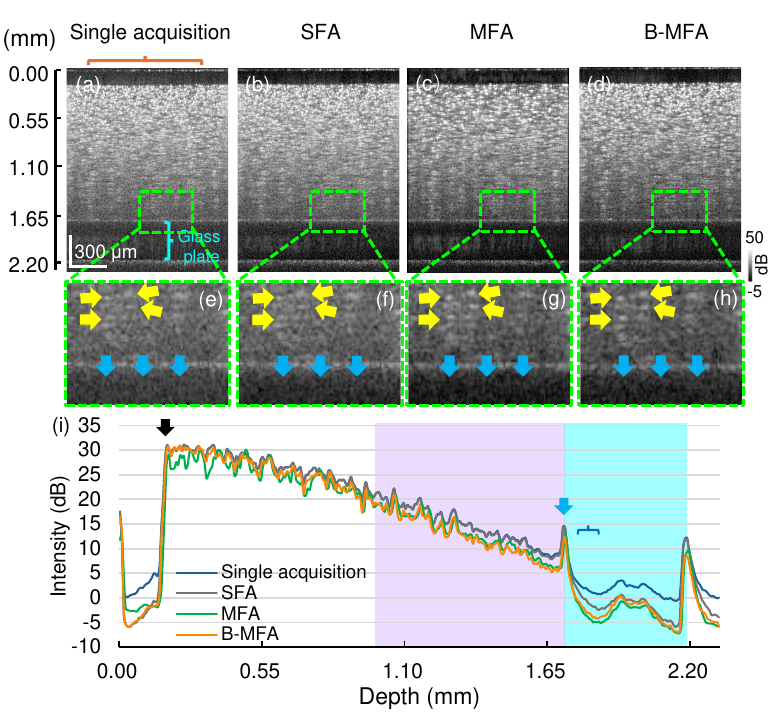}
	\caption{%
		Cross-sectional images obtained by the single acquisition (a), SFA (b), MFA (c), and B-MFA methods (d).
		The scale bar indicates 300 \um. 
		The regions indicated by the green dashed boxes are enlarged in the second row (e--h).
		In the superficial regions in the glass plates (blue arrows), the MFA and B-MFA images have lower signal intensities than the other images because of the reduction of the MS signal.
		(i) Averaged intensity depth profiles of each image, which are averaged using A-lines in the orange brace in (a), where zero dB refers to the noise floor (the signal in the air region) in the single acquisition image.
		The black and blue arrows indicate the surfaces of the scattering region and the glass plate, respectively.
		In the deep region of the scattering sample (purple shading) and the superficial area of the glass plate (light-blue shading), the MFA and B-MFA show lower signal intensity than the other methods because of the reduction in MS.
	}
	\label{fig:res-phantom}
\end{figure}
Figure \ref{fig:res-phantom} shows the intensity cross-sectional images of the scattering phantom for the single acquisition, SFA, MFA, and  B-MFA methods from left to right.
The images in the second row are magnified views of the deep regions, indicated in the images in the first row.
In the shallow regions of the sample, the standard MFA image has fewer particles than the other images [Fig.\@ \ref{fig:res-phantom}(c)].
This difference can be attributed to the fact that only the standard MFA uses 2-D computational refocusing, whereas the others use 1-D refocusing.
This is discussed in detail in Section \ref{sec:limitOf1D}.
In the deep regions (indicated by the green boxes), the B-MFA and MFA images [Fig.\@ \ref{fig:res-phantom}(h) and (g), respectively] show particles with higher contrast than the single acquisition and SFA images [Fig.\@ \ref{fig:res-phantom}(e) and (f)], as indicated by the yellow arrows.
Inside the glass plate region, the B-MFA and MFA images [Fig.\@ \ref{fig:res-phantom}(h) and (g), respectively] have the lowest noise intensity (indicated by the blue arrows), whereas the single acquisition and SFA images [Fig.\@ \ref{fig:res-phantom}(e) and (f)] have the highest and intermediate noise intensities, respectively.
This may indicate that standard OCT measurement noise has been mitigated in the SFA image because of the complex averaging, while both the measurement noise and MS signal are mitigated in the MFA and B-MFA images.

This noise and MS suppression are more clearly and quantitatively shown in the averaged intensity depth profiles in Fig.\@ \ref{fig:res-phantom}(i).
Here, the central 216 A-lines [indicated by the orange brace in Fig.\@ \ref{fig:res-phantom}(a)] were averaged.
The black arrow indicates the surface of the scattering layer.
In the superficial region of the scattering layer, the single acquisition (blue), SFA (gray) and B-MFA (orange) curves are similar to each other.
In the deeper regions (the purple-background region in the plot), the B-MFA and MFA curves show intensities that are lower than those of the single acquisition and SFA curves by around 2 to 3 dB.
The glass plate region is indicated by light blue background in the plot.
At the surface of the glass plate (blue arrow), the single acquisition and SFA curves show higher intensity than those of MFA and B-MFA due to overlapping MS signals.
Near the superficial depth in the glass (indicated by the blue brace), the single acquisition image (blue line) yields the highest intensity.
The SFA (gray line) shows lower intensity than the single acquisition (2.02-dB reduction in average).
This could be attributed to the reduction in measurement noise caused by the complex averaging.
The B-MFA (orange line) shows a further reduction (2.07 dB with respect to SFA and 4.09 dB with respect to the single acquisition in average).
This may indicate additional suppression of the MS signal.
The MFA (green line) shows the lowest signal values, i.e., the best reductions in measurement noise and MS-signal suppression (0.58 dB with respect to B-MFA and 4.66 dB with respect to single acquisition in average).

\subsection{\Invivo small fish sample}
\label{sec:resFish}
\begin{figure}
	\centering\includegraphics{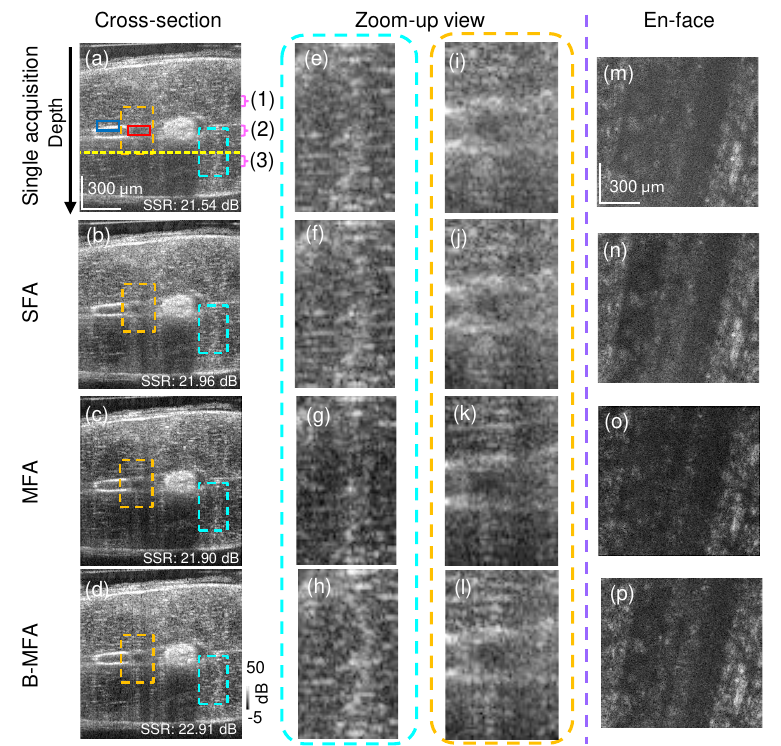}
	\caption{%
		The intensity cross-sections and \enfaceh images of a small fish sample including
		single acquisition (a, m), SFA (b, n), MFA (c, o), and B-MFA (d, p) images.
		The yellow dashed line indicates the depth of the \enfaceh images.
		(e--h) and (i--l) are the enlarged images of the orange and blue boxed regions, respectively.
		The dark-blue and red boxes indicate the regions used to compute the signal-to-signal ratio (SSR).
		The braces, labeled (1), (2), and (3), represent the depth regions used to compute the sharpness metrics.
		The scale bar indicates 300 \um.
		Cross-sectional images of other fish are shown in the Supplementary Material (Fig.\@ \ref{fig:SupDay3Bleomycin1}-\ref{fig:SupDay3Bleomycin3}S).
	}
	\label{fig:fishImage}
\end{figure}
Figure \ref{fig:fishImage} shows the intensity cross-sections of one of the ten fish samples.
The images are, from top to bottom, single acquisition, SFA, MFA, and B-MFA images, respectively.
Enlarged images of the blue and orange regions are shown to the right of the cross-sectional images.
Several anatomic features, such as a layered hyperscattering structure in the muscle region (enlarged in the blue dashed boxes) and a dark region surrounded by hyperscattering, which may indicate a notochord (enlarged in orange dashed boxes) are visible in all images, but these features are most clearly visualized in the B-MFA image. 

Figure \ref{fig:fishImage}(m--p) shows the \enfaceh images at a deep depth indicated by the yellow dashed line in Fig.\@ \ref{fig:fishImage}(a).
The B-MFA image [Fig.\@ \ref{fig:fishImage}(p)] exhibits a lower signal intensity but higher contrast than the single acquisition and SFA images [Fig.\@ \ref{fig:fishImage}(m) and (n), respectively].
These findings may suggest that the B-MFA method reduces the MS signal and improves the image contrast.
We also notice that the \enfaceh MFA image [Fig.\@ \ref{fig:fishImage}(o)] shows reduced signal intensity but does not show improved contrast.
We suspect that this reduced intensity in the MFA image is not fully due to the MS reduction, but is also due to signal washout caused by the motion of the sample.
Details are discussed in Section \ref{sec:signalReduction}.

It might be noteworthy that, since lateral motion is not observed in the \enfaceh images, we reasonably assume that the cross-sectional images in Fig.\@ \ref{fig:fishImage} were taken at the same anatomical position for all methods. 
In addition, we can consider that ``refocus artifact,'' which is discussed in detail later in Section \ref{sec:limitOf1D}, is not remarkable according  to these \enfaceh images.

The intensity cross-sectional image of the other nine fish are presented in the Supplementary Material (Fig.\@ S1-S3).

\begin{table}
	\caption{%
		Information entropy of the \enfaceh slab projections of the small fish.
		The depths of (1)--(3) correspond to the depths indicated in Fig.\@ \ref{fig:fishImage}(a) (braces).
		For all depths, B-MFA shows the smallest entropy, which indicates the sharpest image.
	}
\label{tab:zebrafish}
\centering
\begin{tabular}{c|c|c|c|c}
\multirow{2}*{Depth (mm)} & \multicolumn{3}{c}{Information entropy}\\
\cline{2-5}
& \thead{Single \\ acquisition} & \thead {SFA} & \thead{Standard \\ MFA} & \thead{ B-MFA }\\
\hline
(1) 1.05-1.12 & 4.62 & 4.60 & 4.57 & 4.56\\ 
\cline{1-5}
(2) 1.41-1.48 & 4.18 & 4.16 & 4.08 & 4.06\\
\cline{1-5}
(3) 1.77-1.85 & 4.52 & 4.42 & 4.39 & 4.29\\
\end{tabular}
\end{table}
\begin{figure}
	\centering\includegraphics{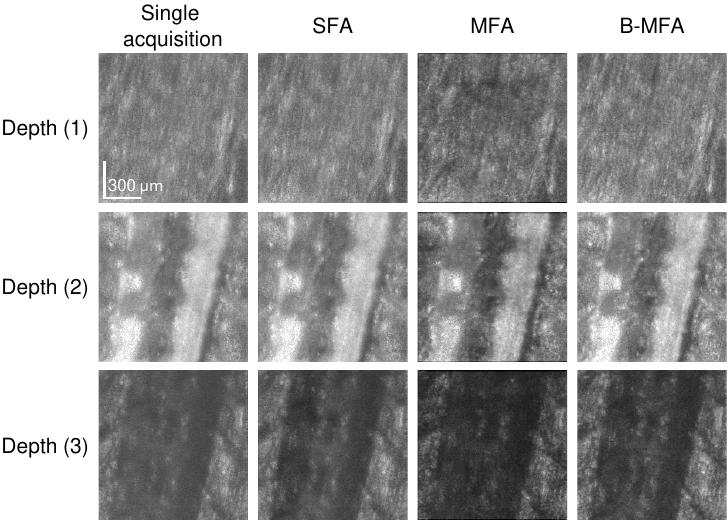}
	\caption{%
		The \enfaceh projection images used to compute the information entropies presented in Table \ref{tab:zebrafish}. 
		It can be found that the information entropies correspond to the observational image contrasts.
	}
	\label{fig:IEfigure}
\end{figure}
To quantitatively compare the sharpness of the images, we computed the information entropy of the \enfaceh slab projections.
The projections were computed at three depth regions indicated by the pink braces in Fig.\@ \ref{fig:fishImage}(a).
The computed information entropy values are summarized in Table \ref{tab:zebrafish}, where the depths are relative to the zero-delay depth.
At all three depth regions, the B-MFA shows the smallest information entropy. 
Namely, B-MFA provides the sharpest image of the four methods.
For reference, the \enfaceh projection images used to compute the information entropy are summarized in Fig.\@ \ref{fig:IEfigure}. 
These images may provide an intuitive understanding of the correspondence between the information entropy values and image sharpness.

\begin{figure}
	\centering\includegraphics{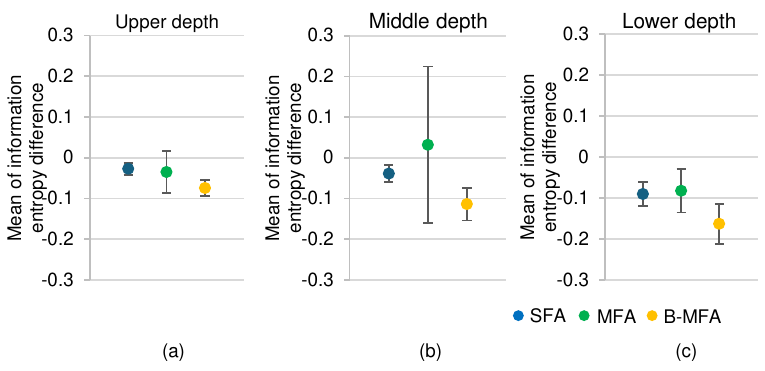}
	\caption{%
		Mean information entropy differences of \enfaceh slab projections of ten fishes are presented. 
		Here, the information entropy difference is the difference of information entropy from that of the single acquisition image and is computed for each fish. 
		Error bars indicate the standard deviation among the ten fishes. 
		The upper, middle, and lower depths were selected for each fish to cover the same structures as in the corresponding three depths of the first fish [brackets in Fig.\@ \ref{fig:fishImage}(a)]. 
		SFA and B-MFA exhibit reasonably small standard deviations, and B-MFA showed noteworthy reductions in mean information entropy compared to SFA and MFA.
	}
	\label{fig:IEplots}
\end{figure}
We performed the same information entropy analysis on the other nine fishes as summarized in Fig.\@ \ref{fig:IEplots}. 
Here, the three depth regions (i.e., upper, middle, and lower) of each fish were selected to cover the same structures as the case of the first fish. 
Since the information entropies may vary among the fishes due to the particular structures of each fish, we computed the information entropy difference from the single acquisition for each fish. 
In the figure, small circles indicate the means among the ten fishes, i.e., the additional nine fish and the first fish presented in Table \ref{tab:zebrafish}, while the whiskers indicate the standard deviation among the ten fishes.

Although MFA shows large standard deviations, SFA and B-MFA show reasonably small standard deviations. 
In comparison to the mean information entropies of SFA and MFA, B-MFA showed smaller information entropies, indicating higher image sharpness. 
By considering the relatively small standard deviations of B-MFA to the mean information-entropy reductions, the reductions (i.e., the sharpness enhancements) are not trivial.

To quantitatively evaluate the image contrast, we computed the signal-to-signal ratio (SSR).
SSR is defined as the mean intensity ratio between two manually selected ROIs, where one ROI was chosen to include a high-scattering-intensity structure [dark blue box in Fig.\@ \ref{fig:fishImage}(a)] and the other was selected in a low scattering region [red box in Fig.\@ \ref{fig:fishImage}(a)] in a cross-sectional image.
The size of each ROI was 35 pixels $\times$ 16 pixels (205.1 \um $\times$ 115.8 \um).
Examples of the SSRs are shown in Fig.\@ \ref{fig:fishImage}(a--d).

The SSRs of the SFA, B-MFA, and MFA images were compared with that of the single acquisition image by computing the SSR enhancement (SSRE), which is defined as the difference in SSR from that of the single acquisition image.
For the images presented in Fig.\@ \ref{fig:fishImage}, the SSREs were 0.42 dB (SFA), 1.37 dB (B-MFA), and 0.36 dB (MFA).
Namely, of the four images, the B-MFA image provides the largest SSRE.  

\begin{figure}
	\centering\includegraphics{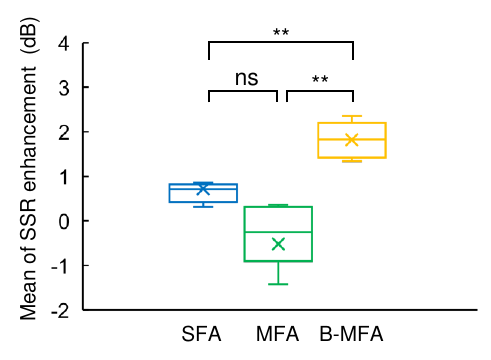}
	\caption{%
		Box plot of the SSR enhancement (SSRE) with respect to the single acquisition images of ten fish samples .
		The horizontal labels represent the methods.
		The top and bottom of each box indicate the upper and lower quartiles, respectively.
		The center lines indicate the medians.
		Paired t-tests revealed that B-MFA shows a significantly higher SSER than SFA and MFA (p =  0.0008 and 0.0009, respectively).
		** represents the statistical significance of p < 0.01, and ns stands for ``non-significant.''
	}
	\label{fig:fishPlot}
\end{figure}
The SSREs of all ten fish were computed using ROIs that include the same anatomical structures for all fish and plotted in Fig.\@ \ref{fig:fishPlot}.
The B-MFA achieved the best mean SSRE of 1.82 dB, which is significantly larger than that of SFA (SSRE = 0.72 dB, p = 0.0008) and MFA (SSRE = -0.52 dB, p = 0.0009).
The statistical comparison was done using paired t-tests.
It is noteworthy that although the mean SSRE of B-MFA was only 1.82 dB (i.e., $\times$ 1.52), the images in Figs.\@ \ref{fig:fishImage} and S1 to S3 demonstrated evidently superior observational contrast of B-MFA images compared to the single acquisition images.

We note that the MFA method showed the largest interquartile range.
This is because the MFA is more susceptible to sample motion, and the motions of the \invivo fish samples highly varied from case to case.
This susceptibility of MFA to the sample motion indicates that the B-MFA might be the best method for \invivo small fish samples.


\section{Discussion}
\subsection{Scan protocol optimization}
\label{sec:optimization}
\begin{table}
	\centering
	\begin{tabular}{c|c|c|c}
		\thead{Measurement \\ protocol} & \thead{$\Delta z$ } & \thead {N} & \thead{$\D = (\mathrm{N}-1) \times \Delta z$} \\
		\hline
		1 & 1 DOF (0.36 mm) & 1, 2, 3, 4 & 0, 0.36, 0.72, 1.08\\ 
		\hline
		2 & 1/2 DOF (0.18 mm) & 1, 2, 3, $\cdots$, 7 & 0, 0.18, 0.36, $\cdots$, 1.08\\ 
		\hline
		3 & 1/3 DOF (0.12 mm) & 1, 2, 3, $\cdots$, 10 & 0, 0.12, 0.24, $\cdots$, 1.08\\ 
		\hline
		4 & 1/4 DOF (0.09 mm) & 1, 2, 3, $\cdots$, 13 & 0, 0.09, 0.18, $\cdots$, 1.08\\ 
		\hline
		5 & 1/6 DOF (0.06 mm) & 1, 2, 3, $\cdots$, 15 & 0, 0.06, 0.12, $\cdots$, 0.84\\ 
	\end{tabular}
	\caption{%
		Summary of B-MFA measurement protocols used for the scan-protocol optimization.
		We used five focus shifting steps ($\Delta z$) and several values for the number of frames to be averaged (N).
		The total focus shifting distance (D) was defined from $\Delta z$ and N.
	}
	\label{tab:optimizationProtocol}
		
\end{table}
To determine the optimal measurement protocol for B-MFA, five focus shifting steps ($\Delta z$) and several values for the number of total frames per B-scan (N) were examined.
The results for the focus shifting step, total frame number, and total focus shifting distance $\mathrm{D} = (\mathrm{N}-1) \times \Delta z$ are summarized in Table \ref{tab:optimizationProtocol}.
A scattering phantom (Section \ref{sec:samples}) was measured using all protocols.

\begin{figure}
	\centering\includegraphics{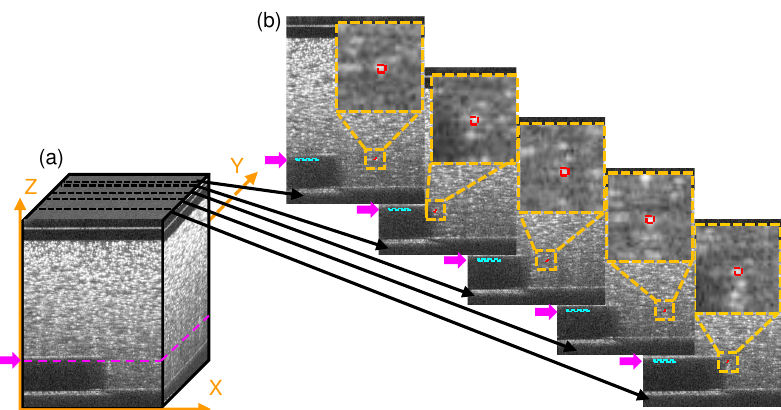}
	\caption{
		Schematic illustration for the scatterer selection.
        Five scatterers (red boxes) and a superficial region in the glass plate (dashed light-blue boxes) were selected to compute the signal-to-background ratio (SBR).
        The former (red boxes) is used to compute the signal level, whereas the latter (light-blue boxes) is used to compute the background level.
	}
	\label{fig:optSchematic}
\end{figure}
To evaluate the image contrast, the signal-to-background ratio (SBR) was computed.
Here the ``signal'' was defined as the mean signal intensity of five manually selected particles in a scattering region of the phantom.
The particles were selected at a depth 10 pixels below the top surface depth of the glass plate as schematically indicated by the pink dashed line in Fig.\@ \ref{fig:optSchematic}(a), and each of the five particles were selected from different cross-sectional image in a volume.
Each scatterer was cropped by a 3 $\times$ 3 pixels (17.6 $\times$ 21.7 \um) window in a cross-sectional image, as indicated by the red boxes in Fig.\@ \ref{fig:optSchematic}(b), and the mean intensity of this window was used as the intensity of that scatterer.
The background was defined as the mean signal intensity of an ROI located at the same depth as the particles but in the glass plate, i.e., a region without scattering.
The ROI extends over 60 pixel $\times$ 3 pixel (351.5 \um $\times$ 21.7 \um, lateral times depth), and is indicated by the dashed light blue box in Fig.\@ \ref{fig:optSchematic}(b).
Because there should be no scattering in the glass, the background intensity is used as a measure of the MS signal.

\begin{figure}
	\centering\includegraphics{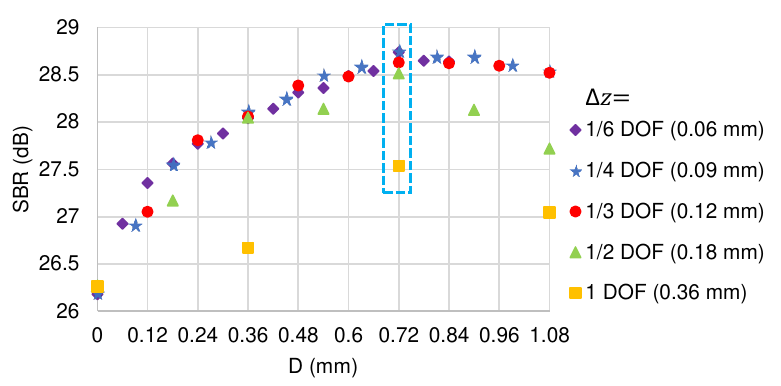}
	\caption{%
		SBRs obtained from several B-MFA measurement protocols.
		Each symbol in the plot represents a focus shifting step ($\Delta z$), and the horizontal axis represents the total focus shifting distance (D).
		The highest SBR was obtained with D = 0.72 mm ($2\times\mathrm{DOF}$).
	}
	\label{fig:optimizationRes}
\end{figure}

The SBR of each protocol is shown in Fig.\@ \ref{fig:optimizationRes} where each symbol corresponds to each focus shifting step $\Delta z$.
The plot demonstrates that the SBR is highest at $\D = 0.72$ mm, which is twice the DOF.
At this $\D$, the SBRs are 28.74, 28.76, 28.63, 28.52, and 27.54 dB for $\Delta z$ = 1/6, 1/4, 1/3, 1/2, and 1 $\times$ DOF, respectively.
Here, the numbers of complex averaged frames (N) of each protocol are 13, 9, 7, 5, and 3, respectively.
Among these five protocols, the first four give similarly high SBR, and they can be the candidates for the optimal protocol. 

Because B-MFA is used for \invivo measurements, a shorter acquisition time is preferable. 
Therefore, we have used the protocol of ($\Delta z$ = 1/2 DOF, N = 5, $\D = 2$ DOF) for the measurements shown in the Section \ref{sec:allresults}.

Note that this protocol was selected to best suit small fish samples. 
For other types of samples, it could be worth optimizing the protocol again to adapt it for the new samples.

\subsection{Phase stability requirements of MFA and B-MFA}
\label{sec:MFAvsBMFA}
The previously proposed MFA method used a 2-D computational refocusing \cite{zhu2022computational}, in which the complex \enfaceh OCT signal at a depth is two-dimensionally Fourier-transformed and 2-D quadratic phase is applied in the spatial-frequency domain.
Hence, the phase of the OCT signal should be stable over the volume or at least over several frames that cover an area larger than the lateral resolution. 

In contrast, B-MFA uses 1-D computational refocusing as described in Section \ref{sec:1dRefocusing}.
Hence, the phase stability is required only within a frame.

Because the sample motion can cause significant phase error, the lower requirement for the phase stability of the B-MFA is an important advantage for \invivo measurement.


\subsection{Refocus artifact of B-MFA and conventional MFA in the phantom result}
\label{sec:limitOf1D}

\begin{figure}
	\centering\includegraphics{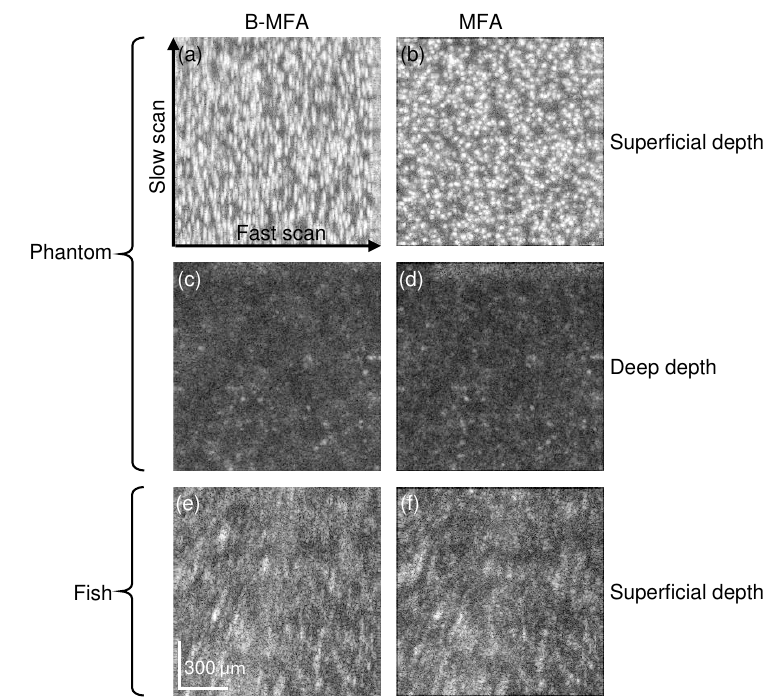}
	\caption{%
		Refocus artifacts in \enfaceh images of the phantom (first two rows) and a small fish sample (the bottom).
        The left images were obtained by the B-MFA method, whereas the right images were obtained by the MFA method.
        Because the 1-D computational refocusing used by B-MFA refocuses the imaging only along the fast-scan direction (horizontal direction of these images), the structures in the B-MFA images are elongated along the slow-scan direction (vertical direction in the image).
        In contrast, MFA image does not show this effect because it uses 2-D refocusing.
        In the B-MFA images, the elongation is less pronounced in the deep depths of the phantom (c), which is because the physical focus was close to this depth, and hence, the structure was sharp even without refocusing.
	} 
	\label{fig:1Dvs2Drefocus}
\end{figure}
Comparing the B-MFA and MFA cross-sectional images [Fig.\@ \ref{fig:res-phantom}(d) and (c), respectively] at the superficial region scattering layer, we notice that the B-MFA image exhibits more scattering particles than the MFA image.
This difference in the numbers of particle can be attributed to the difference in the 1-D and 2-D computational refocusing.
Because 1-D refocusing refocuses the image only along the fast-scan direction, the refocused particle signal must be elongated along the slow-scan (vertical) direction, as shown in the \enfaceh slice of the same data [Fig.\@ \ref{fig:1Dvs2Drefocus}(a)].
Hence, the images of the scatterers that are not really in a particular B-scan smear into the B-scan, and this causes an artifactual increase in the number of scatterers in that B-scan.
On the other hand, the 2-D refocusing used in MFA isotropically refocuses the image as shown in the corresponding \enfaceh slice [Fig.\@ \ref{fig:1Dvs2Drefocus}(b)].
Hence, the artifactual increase of the scatterer does not occur.

A similar artifact also can be seen in the small fish image shown in Fig.\@ \ref{fig:1Dvs2Drefocus}(e) and (f), although it is less evident than in the phantom case, because the tissue microstructure is aligned roughly along the slow-scan direction.

In the other depth of the phantom, which is around 1.5 mm from the surface, the elongation artifact is negligible [Fig.\@ \ref{fig:1Dvs2Drefocus}(c) and (d)] because the physical focus is located near this depth.

To solve this problem in the future, we may be able to adopt a computational refocusing method\cite{RuizLopera2020OL, RuizLopera2023OL} for the MFA or B-MFA method that is less susceptible to phase instability.

\subsection{Signal reduction of the MFA in the \invivo result}
\label{sec:signalReduction}
In the \invivo fish measurements, the MFA image [the third row of Fig.\@ \ref{fig:fishImage}] showed lower signal intensity than the B-MFA image [the fourth row of Fig.\@ \ref{fig:fishImage}].
In addition, the SSRE of MFA is smaller than that of B-MFA for the \invivo measurement (ninth paragraph of Section \ref{sec:resFish}).
This can be attributed to the signal washout caused by the sample motion, which reduces not only the MS signal but also the SS signal.
These findings also emphasize the advantage of the B-MFA method for \invivo imaging over the MFA method.

\subsection{Heartbeat and respiration of fish}
\label{sec:motionOffish}
The heart and respiratory rates of medaka fish are approximately 2.3 Hz and 4.9 Hz, respectively \cite{watanabe2022acquirement}. 
These correspond to heart and respiratory periods of 0.43 s and 0.20 s, respectively. 
On the other hand, the B-MFA acquisition of a single cross-section (i.e., acquisition time for five consecutive frames) is 67.1 ms (Section \ref{sec:measurementProtocol}), which is 6.5 times and 3 times shorter than the heart and respiratory periods, respectively. 
This fact further supports the feasibility of \invivo application of B-MFA.

However, this comparison between the B-MFA acquisition time and the heart and respiratory periods also suggests that some B-MFA cross-sections may have been affected by sample motion, causing signal reduction. 
Overcoming this issue remains a future challenge.

\subsection{Extensions of B-MFA-based OCT}
\label{sec:estensionsOfbmfa}
\subsubsection{Functional OCT and B-MFA}
\label{sec:bmfaCompatibility}
One limitation of B-MFA is its incompatibility with functional OCT imaging techniques based on fast sequential acquisition, including Doppler OCT \cite{White2003, Leitgeb2003, Baumann2011}, OCT angiography \cite{Makita2006, Jia2012}, and certain types of dynamic OCT (DOCT) \cite{Apelian2016, Muenter2020, ElSadek2020}. 
This limitation stems from B-MFA's requirement for the sequential acquisition of a few frames followed by complex averaging. 
As a result, the effective frame rate of B-MFA cross-sectional images is reduced, and the complex averaging process can wash out not only the MS signal components but also the dynamic signal components.

On the other hand, certain types of DOCT methods employ a relatively slow time sequence of OCT images. 
For instance, volumetric logarithmic intensity variance (LIV) and volumetric OCT correlation decay speed (OCDS) methods use a sequence of 32 OCT frames with a frame interval of 204.8 ms \cite{abd2021oct}. 
Given that the acquisition time of a cross-sectional B-MFA image (i.e., the acquisition time of a set of five frames) is only 67.1 ms (as discussed in Section \ref{sec:measurementProtocol}), B-MFA is theoretically compatible with LIV- and OCDS-based DOCT.
In this scenario, relatively fast dynamics may be washed out, but slow dynamics can still be detected to some extent. Therefore, experimental validation of B-MFA-based dynamic OCT could be a promising avenue for future research.

B-MFA may also be compatible with static functional OCT, such as polarization-sensitive OCT. 
Although we used the JM-OCT system in this study without exploiting its polarization sensitivity, a minor modification of the signal processing may enable polarization-sensitive B-MFA imaging. 
It is worth noting that Lida Zhu \etal have demonstrated MFA-based polarization \exvivo imaging and showed that MFA can reduce polarization artifacts \cite{zhu2024polarization}. 
Therefore, polarization-sensitive B-MFA may mitigate the polarization artifact in \invivo imaging.

\subsubsection{Extension of measurement filed}
\label{sec:FovOfbmfa}
For certain \invivo applications, wide-lateral-field imaging is of significant interest.
However, B-MFA necessitates lateral oversampling due to its utilization of computational refocusing, and it may limit the field size. 
For instance, in the present study, the lateral B-scan area is 1.5 mm and is covered with 256 sampling points.

One potential solution is the adoption of high-speed OCT systems. 
While the current study employs an OCT system with an A-line rate of 50,000 A-lines/s, ultra-high-speed OCT systems capable of A-line rates such as 3,280,000 A-lines/s \cite{Goeb2022} have been demonstrated, recently. 
By utilizing such a system, for example, the lateral size of the B-scan could be extended by up to 65.6 times (i.e., up to 98.4 mm) without compromising the sampling density and without increasing the measurement time in principle.


\section{Conclusion}
We proposed the B-MFA method, which suppresses MS signals in \invivo imaging.
The method was validated using phantom and \invivo small fish measurements.
The subjective observation of the images and objective evaluation of SSRE showed that the B-MFA method improves the image contrast by reducing the MS signals.
In addition, the B-MFA showed superior performance than MFA for \invivo measurements.

Here, we conclude that the B-MFA method can reduce noise caused by the MS signal in OCT images and is a better option for \invivo measurement than our previously proposed MFA method.

\appendix
\section*{Appendix}
\section{Sample-surface detection}
\label{sec:segmentation}
The details of the sample surface detection used in Section \ref{sec:complexAverage} are as follows.
We first manually select the rough depth region in which the sample surface is searched, which is typically around 90 pixels (around 652 \um in air).
Then, we compute the first order derivative of linear OCT intensity along the depth, where the derivative is defined as the difference between two neighboring pixels.
The surface is defined at the top-most depth where the derivative is larger than a predefined threshold.
The threshold is 50 in our particular case, but this value is in arbitrary unit and may differ in various OCT systems.

\section*{Funding}
Core Research for Evolutional Science and Technology (JPMJCR2105); 
Japan Society for the Promotion of Science (21H01836, 22K04962, 22KF0058);
China Scholarship Council (201908130130).

\section*{Acknowledgment}
Please see \email{\authormark{}https://optics.bk.tsukuba.ac.jp/COG/.}

\section*{Disclosures}
Y. Zhu, L. Zhu, Lim, Makita, Guo, Yasuno: Sky Technology (F), Nikon (F), Kao Corp. (F), Topcon (F), Panasonic (F), Santec (F). 
L. Zhu is currently employed by Santec.

\section* {Data Availability} 
Data underlying the results presented in this paper are not publicly available at this time but may be obtained from the authors upon reasonable request. 

\section*{Supplemental document}
See Supplement 1 for supporting content.

\bibliography{reference}

\pagebreak
\title{Supplementary Material}
\setcounter{figure}{0}
\renewcommand\thefigure{S\arabic{figure}}   
\setcounter{table}{0}
\renewcommand\thetable{S\arabic{table}}   
\setcounter{section}{0}
\renewcommand\thesection{S\arabic{section}}  

\begin{abstract}
This file supplements Section \ref{sec:resFish} by showing the intensity cross-sectional images of the other medaka fish samples (samples 2 to 10) measured for validation of MS reduction by the B-scan-wise-multi-focus averaging (B-MFA) method. 
All of the results show that several anatomic features, such as a hyper-scattering layer structure and the boundary of hollow structure are better visible in the B-MFA image than single acquisition, SFA, and conventional MFA. 
\end{abstract}

\begin{figure}[ht]
    \centering\includegraphics{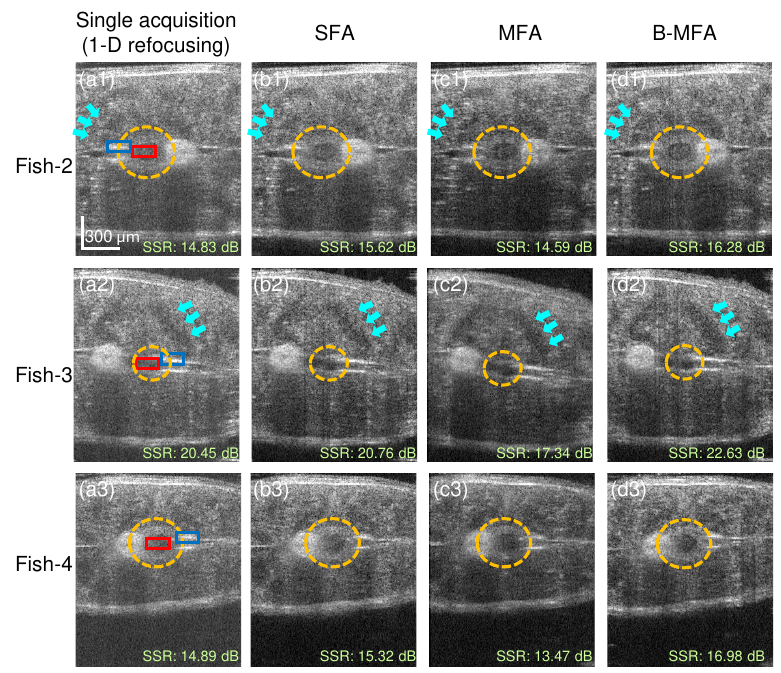}
    \caption{
        The intensity cross-sectional mages of the medaka (samples 2--4).
		 (a1-a3) single acquisition, (b1-b3) SFA, (c1-c3) conventional MFA, (d1-d3) B-MFA image. 
		 The blue and red boxes indicate the region used to compute the signal-to-signal ratio (SSR). 
    }
    \label{fig:SupDay3Bleomycin1}
\end{figure}	
	\begin{figure}
		\centering\includegraphics{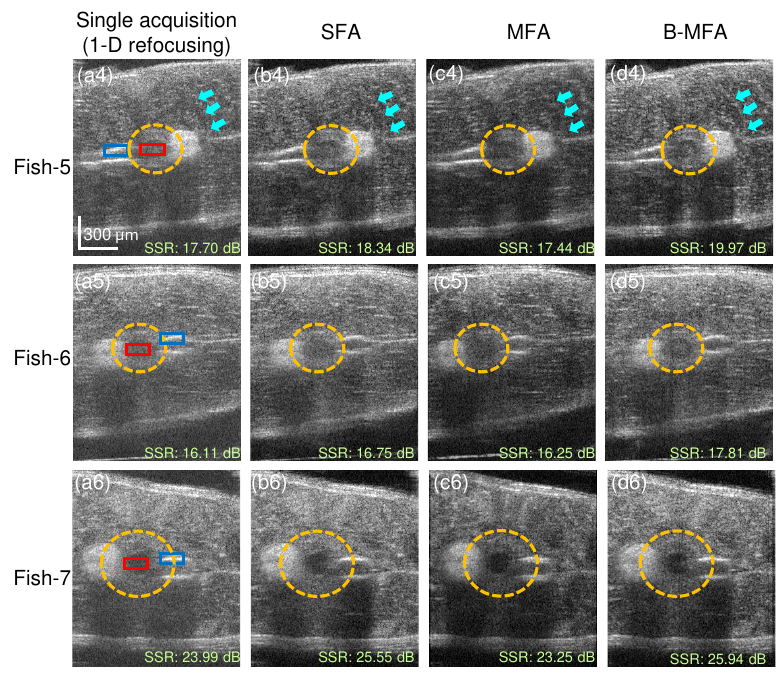}
		\caption{%
			The intensity cross-sectional image of medaka (samples 5--7).
			The image types and their order are identical to those of Fig.\@ \ref{fig:SupDay3Bleomycin1}.  
		}
		\label{fig:SupDay3Bleomycin2}
	\end{figure}	
 	\begin{figure}
		\centering\includegraphics{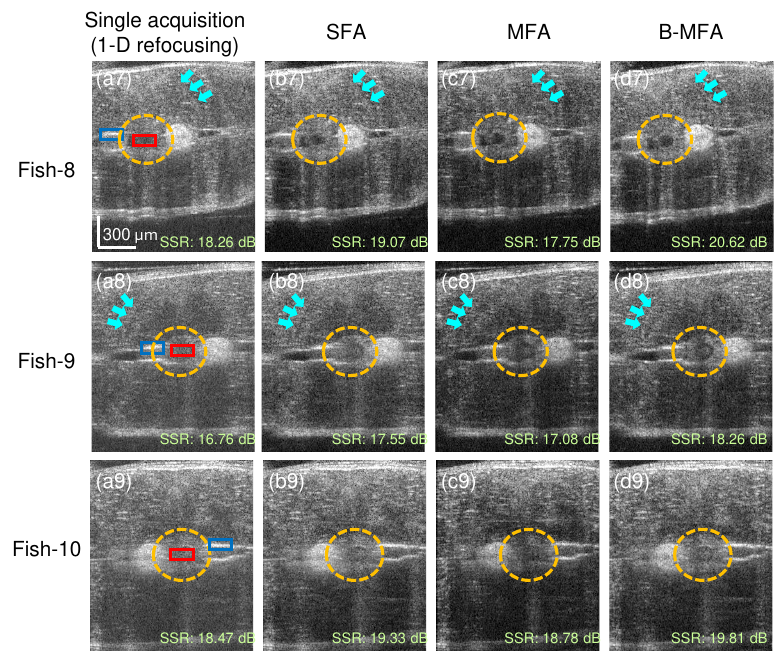}
		\caption{%
			The intensity cross-sectional image of medaka (samples 8--10).
			The image types and their order are identical to those of Fig.\@ \ref{fig:SupDay3Bleomycin1}. 
		}
		\label{fig:SupDay3Bleomycin3}
	\end{figure}	
\
\end{document}